\documentclass[preprint, dvips]{imsart}

\RequirePackage[OT1]{fontenc}
\RequirePackage{amsthm,amsmath}
\RequirePackage[numbers]{natbib}
\RequirePackage[colorlinks,citecolor=blue,urlcolor=blue]{hyperref}
\RequirePackage{graphicx, graphics, epsfig, amsfonts, amssymb}
\RequirePackage[parfill]{parskip}
\RequirePackage[small,compact]{titlesec}

\startlocaldefs
\numberwithin{equation}{section}
\theoremstyle{plain}
\endlocaldefs

\begin{document}

\begin{frontmatter}
\title{Photon-Assisted Process and High-Harmonic Dynamic Localization in Graphene Nanoribbons}
\runtitle{Photon Assisted Peaks in graphene nanoribbons}

\begin{aug}
\author{\fnms{Rabiu} \snm{Musah}\thanksref{T1}\ead[label=e1]{rabpeace10gh@gmail.com}}
\address{Department of Applied Physics, Faculty of Applied Sciences, University\\
for Development Studies, Navrongo Campus, Ghana.\\
\printead{e1}}

\author{\fnms{Samuel Y.} \snm{Mensah}\thanksref{T1}\ead[label=e2]{profsymensah@yahoo.co.uk}}
\and
\author{\fnms{Suleimana S.} \snm{Abukari}\thanksref{T1}\ead[label=e3]{asseidu75@yahoo.com}}
\address{Department of Physics, Laser and Fiber Optics Center, University \\ 
of Cape Coast, Cape Coast, Ghana.\\
\printead{e2,e3}}

\thankstext{T1}{Also, Nanoscience Research Group, University of Cape Coast, Ghana}
\runauthor{Rabiu Musah et al.}
\end{aug}

\begin{abstract}
We used a complete tight-binding band structure of graphene nanoribbon to obtain, for the first time, analytical techniques for observing photon assisted transport, and dynamic localization of electrons in the graphene nanoribbons. When the ribbons are subject to a multi-frequency dc-ac field, photon assisted replicas show up at rather strong drive force. The strong dependence of the photon peaks on ac amplitudes allow for high-harmonic dynamic oscillations at these amplitudes. We identified regions of positive differential conductivity where a nanoelectronic graphene device may be operated as a small signal amplifier. Our research has also reveal another quantum mechanical phenomenon, fractional photon assisted transport, when the stark factor $r > 1$. 
\end{abstract}

\begin{keyword}
\kwd{Graphene}
\kwd{Current density}
\kwd{Bessel function}
\kwd{Negative differential conductivity}
\end{keyword}

\received{\smonth{2} \syear{2012}}
\end{frontmatter}

\section{Introduction}
Photon assisted processes in systems are quantum phenomenon in the presence of high electric fields in which electrons can absorb or emit one or several photons. These absorption or emission of photons are seen as new conduction channels in the system. The observation of photon assisted process in solid sate materials is central ingredient for terahertz (THz) generations. In semiconductor nanostructures, photon assisted peaks (PAPs) has emerged as a powerful tool for investigating quantum transport phenomenon. PAPs up to 7 THz photons has been reported in \cite{ZeunerS} and can persist up to room temperature. Recent studies have addressed electronic properties of confined graphene structures like dots, nanowires, nanotubes (NTs) and nanoribbons (NRs). In particular, nanoribbons have been suggested as potential candidate for replacing electronic components in future nanoelectronics devices \cite{GeimA, SonJ}. Research have also shown that graphene is a suitable candidate to examine photon assisted processes in Dirac systems \cite{CastroN}. For this reason among others, photon assisted tunneling was studied quite recently in graphene  \cite{IurovA} and also \cite{TrauzettelB}. In NTs, PAPs occurs because the current dynamics in the system can be seen as Bloch oscillations when an ac field is added to dc field but with frequency somewhat below the Bloch frequency associated with the dc field. The large ac amplitude opens up transport channels which can be seen as photon peaks. Regions of positive differential conductivity at certain ranges of the dc field are formed. As a consequence, domainless amplification of THz frequencies can be achieve \cite{AbukariSe}. 

When a superlattice is irradiated with an intense dc-ac field (laser pulse), it can make charged particles localized around the initially incident region, this results in what is called dynamic localization \cite{AWGhosh, XGZhao}. The combined effect of the dc-ac field is crucial and can cause destructive and constructive effects in electron dynamics. In the latter, regions of positive dynamic conductivity are formed. The destructive case is manifested by what the dynamic localization unless the ratio of Bloch frequency to the applied frequency is not an integer and that the argument of a Bessel function whose order is an integer is not a root \cite{WXYan}.

In this paper we investigate the possibility of photon assisted processes (replicas) in GNRs using Boltzmann transport equation in the relaxation time approximation. Motivated by the fact that photon assisted tunneling has recently appeared in \cite{IurovA} where a very simple Dirac spectrum describing only low energy electrons was used,  we explore not the tunneling of Dirac electrons through barriers but the operation regions for possible THz amplification using full tight binding spectrum of graphene. We will show that the I-V characteristics demonstrate regions of positive differential conductivity (PDC) where a graphene nanodevice can be operated as a small signal amplifier. We noted that a dynamical response of the electrons to terahertz radiation can depend on details of scattering processes, this effect could lead to the relaxation towards a thermal equilibrium which we relax here. Again we ignore transient processes, that is we assume that $t >> \tau$ \cite{YuARamanov}.

The rest of the article is organized as follows; we derived the photon assisted current density in section \ref{sec:PAPcurrent}. Using the photon currents, we study the photon replicas and dynamic localization in the presence of monochromatic and biharmonic frequencies in section \ref{sec:PAPeaks}. I-V characteristic graphs are plotted and discussed in details in section \ref{sec:Results} and the paper finally concludes in section \ref{sec:Conclusions} outlying some further investigations.

\section{Photon assisted current density}\label{sec:PAPcurrent}
The current density of electrons in GNR subject to a general multi-frequency field, $E(t) = E_0 + \sum_{j=1}^{n}E_jcos(\omega_j t + \alpha_j)$ is \cite{NotYet}
\begin{equation}
	j = i\sum_{r = 1}^{\infty}j_{0r}\left[ \sum_{n_j,\,\nu_j=-\infty}^{\infty}\prod_{j=1}^nJ_{n_j}(r\beta_j)J_{n_j-\nu_j}(r\beta_j)\frac{e^{i\nu_j\omega_jt + i\nu_j\alpha_j}}{1+i\tau(r\beta_0 + n_j\omega_j)} + c.c\right],\label{eq:jaPAPs}
\end{equation}
where $J_n(x)$ are Bessel functions of $n^{th}$ order. $\beta_j = elE_j/\hbar\omega$, $\beta_0 = elE_0/\hbar$ and $j_0$ is the peak current define as 
\[
	j_{0r} = \frac{2g_sg_ve\gamma_0}{\pi l\hbar}\Delta\theta\sum_{s=1}^{n}r\mathcal{E}_{rs}f_{rs}.
\]
$\Delta\theta = \pi/(\mathcal{N} + 1)$ and $l = \sqrt{3}a/2$ for aGNR, $\Delta\theta = (2 + 1/s)\pi/(\mathcal{N} + 1)$ and $l = a/2$  for zGNR. In order to have more simplified equation that describes the phenomenon we are considering, we adopt the following definitions;
\[
	\beta_0\tau = \frac{E_0}{E_{cr}},\qquad \quad E_{cr} = \frac{\hbar}{el\tau}
\]
\[
	\beta_j = \frac{E_j}{E^*_{j}},\qquad \quad E^*_{j} = \frac{\hbar\omega_j}{el}
\]
with $\omega_j\tau =E^*_j/E_{cr}$. $E^*$ is the electric field at which an electron emits or absorbs a photon when displaced by quasi-lattice period $l$. $E_{cr}$ is however, the critical field at which carriers get their peak velocity. With these choice of variables we make the replacement $	\beta_0\tau + n_j\omega_j\tau \to E_0/E_{cr} + n_jE^*_j/E_{cr}$, so that Eq.\eqref{eq:jaPAPs} becomes
\begin{eqnarray}
	j(t) &=& i\sum_{r = 1}^{\infty}j_{0r} \left[ \sum_{n_j,\,\nu_j=-\infty}^{\infty}\prod_{j=1}^nJ_{n_j}\left(r\frac{E_j}{E^*_j}\right)J_{n_j-\nu_j}\left(r\frac{E_j}{E^*_j}\right) \right]\nonumber\\
			 && \times f\left(r\frac{E_0}{E_{cr}} + n_j\frac{E^*_j}{E_{cr}}\right)e^{i\nu_j\omega_jt + i\nu_j\alpha_j} ,\label{eq:jaPAPs1}
\end{eqnarray}
where $f(x)$ are superposition of weighted photon replicas of  pure dc differential current density. If the above equation is compared to the pure dc current density in reference \cite{NotYet}, we see that the photon peaks are displaced by multiples of $E^*/E_{cr}$ with amplitudes $J_n^2$. Note that if $r=1$ and $n_j=1$, the Bloch frequency, $\omega_B$ ($=elE_0/\hbar$) associated with the static field will coincide with the applied frequency $\omega_j$ and no photon replicas can be seen. Therefore it is a requirement that $r \neq n_j$ and $n_j \neq 1$, if PAPs are to be observed. In the following section, we will consider two cases of Eq.\eqref{eq:jaPAPs1}; the monochromatic and bichromatic cases in the presence of the static electric field.

\section{Large-signal dynamic localization and photon-assisted replicas}
\label{sec:PAPeaks}
\subsection{Graphene nanoribbon in monochromatic field}
When an ac field of frequency, $\omega$ is applied to the GNR, we have $n = 1$, $\omega_{j>1} = 0$ and $\alpha_j=0$ in Eq.\eqref{eq:jaPAPs1} which becomes

\begin{equation}
	j = i\sum_{r = 1}^{\infty}\frac{j_{0r}}{E} \left[ \sum_{n,\,\nu=-\infty}^{\infty}J_{n}\left(r\frac{E}{E^*}\right) J_{n-\nu}\left(r\frac{E}{E^*}\right) f\left(r\frac{E_0}{E_{cr}} + n\frac{E^*}{E_{cr}}\right) \right]\,Ecos(\nu\omega t).\label{eq:jaPAPs2}
\end{equation}
The coefficient of $Ecos(\omega t)$ in Eq.\eqref{eq:jaPAPs3} is an ac part of the drive field. It is seen as a large-signal dynamic conductivity at the derive harmonic frequency $\omega$. The fundamental derive frequency is obtained by setting $\nu=\pm 1$. To simplify the preceding equation further, we take $\nu=0$ to get $cos(\nu\omega t) = 1$, so that

\begin{equation}
	j = i\sum_{r = 1}^{\infty}j_{0r}\left[ \sum_{n=-\infty}^{\infty}J^2_{n}\left(r\frac{E}{E^*}\right) f\left(r\frac{E_0}{E_{cr}} + n\frac{E^*}{E_{cr}}\right) \right].\label{eq:jaPAPs3}
\end{equation}

\subsection{Graphene nanoribbon in bichromatic field}
Now, for an applied field of frequencies, $\omega_1$ and  $\omega_2$, we have $n=2$ and $\omega_{j>2} = 0$. These set $\alpha_{j>1}=0$. One can put $\nu_1 \neq  0$ and $\nu_2 \neq 0$ allowing Eq.\eqref{eq:jaPAPs2} look like

\begin{eqnarray}
	j &=& i\sum_{r = 1}^{\infty}j_{0r}\sum_{\nu_1,\nu_2=-\infty}^{\infty} \sum_{n_1,n_2=-\infty}^{\infty}J_{n_1}\left(r\frac{E_1}{E_1^*}\right)J_{n_1-\nu_1}\left(r\frac{E_1}{E_1^*}\right)J_{n_2}\left(r\frac{E_2}{E_2^*}\right)\nonumber\\
		&& \times J_{n_2-\nu_2}\left(r\frac{E_1}{E_1^*}\right) f\left(r\frac{E_0}{E_{cr}} + n_1\frac{E^*_1}{E_{cr}} + n_2\frac{E^*_2}{E_{cr}}\right) cos( \nu_2\alpha). \label{eq:jaPAPs4}
\end{eqnarray}
We arrived at the last equation by allowing the two frequencies to be commensurable, i.e $|\nu_1|\omega_1 = |\nu_2|\omega_2$, and periodic, i.e with $\omega_2  = \mu\omega_1$. You can see reference \cite{YuARamanov} and references there in for the case of non-commensurable frequencies. 
Further, if both $\nu_1 = 0$ and $\nu_2 = 0$ then a more simplified equation

\begin{equation}
	j = i\sum_{r = 1}^{\infty}j_{0r}\left[ \sum_{n_1,\,n_2=-\infty}^{\infty}J^2_{n_1}\left(r\frac{E_1}{E_1^*}\right)J^2_{n_2}\left(r\frac{E_2}{E_2^*}\right)f\left(r\frac{E_0}{E_{cr}} + n_1\frac{E^*_1}{E_{cr}} + n_2\frac{E^*_2}{E_{cr}}\right) \right].\label{eq:jaPAPs5}
\end{equation}
is obtained. A significant difference between the above two equations is the phase factor $cos(\nu_2\alpha)$. Of course, among other differences, Eq.\eqref{eq:jaPAPs5} will be an inversion of Eq.\eqref{eq:jaPAPs4} for $\alpha \in [\pi/2, 3\pi/2]$. The advantage of the biharmonic field over monoharmonic field is that in the former, new local structures in the I-V characteristics become most pronounce. We shall see this in the following section.

\section{Results and Discussions} \label{sec:Results}
\begin{figure}[thb!]
	\centering{\includegraphics[scale=0.59]{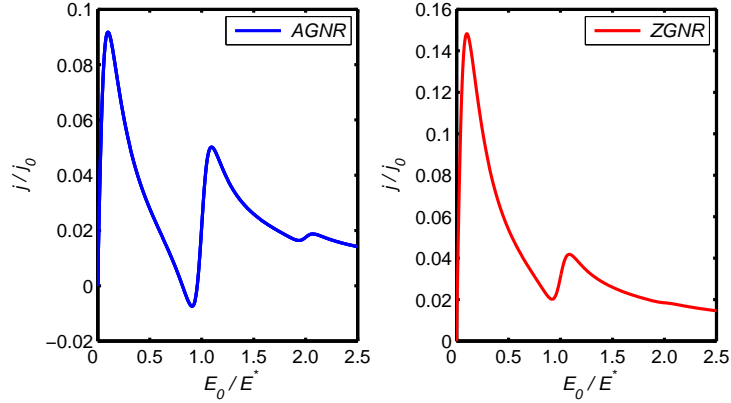}}
	\caption{I-V characteristic showing regions of PDC for (left) armchair and (right) zigzag ribbons. $\omega\tau = 10$, $E = E^*$.}
	\label{fig:azpapss}
\end{figure}

\begin{figure}[thb!]
	\centering{\includegraphics[scale=0.56]{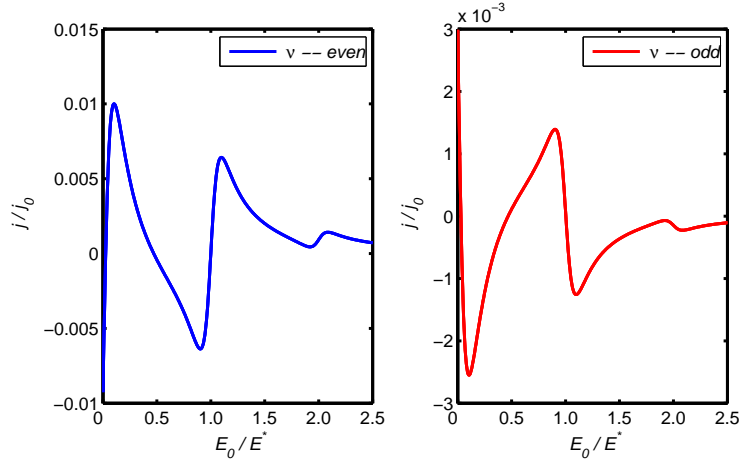}}
	\caption{High-harmonic I-V characteristic showing regions of PDC at (left) even harmonics and (right) odd harmonics for AGNR. $\omega\tau = 10$, $E = 1.6E^*$.}
	\label{fig:azpapssv}
\end{figure}

The graphs in Fig.\ref{fig:azpapss} are $j-E_0$ plot of Eq.\eqref{eq:jaPAPs2} which demonstrate appearance of positive differential conductivity (PDC) regions at high applied frequencies, $\omega\tau \sim 10$. Particularly in AGNR, when the Bloch frequency associated with the bias field falls within the ranges $9.15\tau^{-1} < \omega_B < 10.9\tau^{-1}$, $19.25\tau^{-1} < \omega_B < 20.65\tau^{-1}$ etc., NDC is greatly suppressed and the device can be operated effectively. This means the system is free from space charge instability. Any small signal passing through the ribbon no more suffer from electrical domains and can thus easily get modified. The scenario is the same for ZGNR except that some operation points are hidden at large $E_0/E^*$ values. 
From the figure, because the peaks show up at $E_{cr}/E^* + n > E_0/E^*$ and the minima at $E_0/E^* > n$, where $n$ is the number of peaks, one can deduce a general range of bias field for best device operation to lie within $n < E_0/E^* < n + E_{cr}/E^*$ or 
\[ n\omega\tau < \omega_B\tau < n\omega\tau + 1.
\]
The behavior of Eq.\eqref{eq:jaPAPs3} is shown in Fig.\ref{fig:azpapssv}. PDC effect is observed at high even and odd harmonics. We chose $\nu = 2$, $\nu = 3$ and $cos(\nu\alpha) = 1$. The two graphs are distinct, they are mirror reflections of each other. NDC in AGNR becomes PDC in ZGNR and vice-versa. It is clear from the two curves that depending on the sign of $cos(\alpha)$, the operation ranges in one ribbon can be greater than the other.

\begin{figure}[thb!]
	\centering{\includegraphics[scale=0.54]{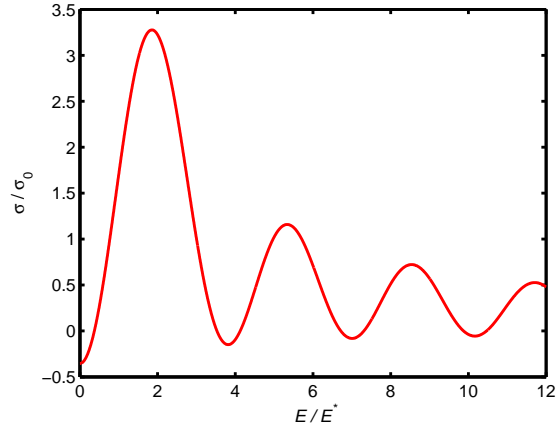}}
	\caption{Dynamic localization states induced by ac-dc fields occur at $E/E^* = 3.8, 7.02, 10.09$. $\nu=0$}
	\label{fig:dynamicloc0}
\end{figure}

To see the existence of dynamic localization in graphene nanoribbons, we plot Eq.\eqref{eq:jaPAPs3} for normalized conductivity versus $E/E^*$ for default parameters $\omega\tau=5$ and $\omega_B\tau = 4$ in Fig.\ref{fig:dynamicloc0}. 

\begin{figure}[thb!]
	\centering{\includegraphics[scale=0.65]{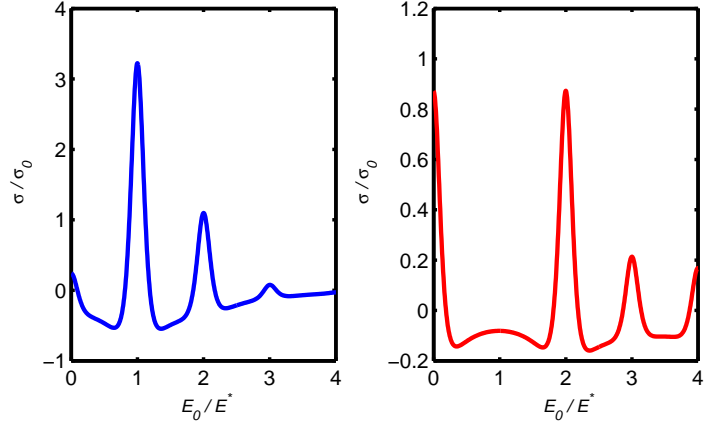}}
	\caption{(blue) Photon assisted peaks for $E/E^* = 2.0$. The peaks occur at $\omega_b = \omega, 2\omega, 3\omega, \ldots$ (red) Dynamic localization at first peak. We chose $E/E^* = 7.0$.}
	\label{fig:dynamicloc_peaks}
\end{figure}

Photon assisted peaks are shown in Fig.\ref{fig:dynamicloc_peaks} (blue) for aGNR. Multi-photon resonances appear when $r\omega_B\tau = n\omega\tau$, for an integer $n$. If $n/r$ is the order of the Bessel functions with root coinciding with one of the $E/E^*$ values in Fig.\ref{fig:dynamicloc0}, a dynamic localization is seen. Depending on how the ratio  $E/E^*$ is chosen, the localized charged carriers can extend the localization beyond its vicinity and thus affecting neighboring centers. This might account for decrease in neighboring peak heights. The suppression at the first photon assisted peak of the red curve is due to the combined destructive effect of ac-dc fields at $E_0 = E^*$ .

\begin{figure}[thb!]
	\centering{\includegraphics[height=2.4in, width=2.8in]{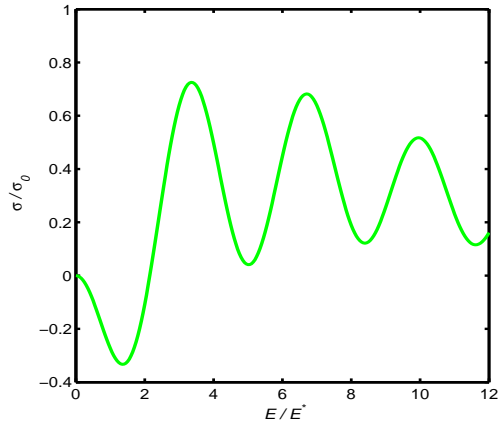}}
	\caption{High dynamic localization states induced by ac-dc fields occur at $E/E^* = 5.05, 8.45, 11.65$. $\nu=2$}
	\label{fig:dynamicloc}
\end{figure}

\begin{figure}[thb!]
	\centering{\includegraphics[scale=0.57]{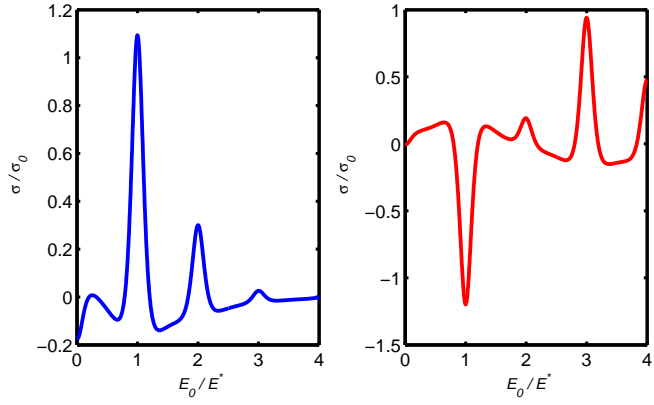}}
	\caption{(blue) Photon assisted peaks for $E/E^* = 2.5$. The peaks occur at $\omega_B = \omega, 2\omega, 3\omega, \ldots$ (red) High-harmonic dynamic localization at first peak. We chose $E/E^* = 5.0$.}
	\label{fig:dynamicloc0_peaks}
\end{figure}

To observe high-harmonic dynamic localization in graphene, we plot the high dynamic conductivity equation with $E/E^*$ for $\mu=2$ in Fig.\ref{fig:dynamicloc}. We obtained the localized states at $E/E^* = 5.05, 8.45, 11.65$. Localized sates appear later in high-harmonic case compared to the normal dynamic localization. This might be an an indication that at high harmonics dynamic localization could disappear giving way for proper direct signal rectification. Fig.\ref{fig:dynamicloc0_peaks} shows photon assisted resonances and high-harmonic dynamic localization at the first photon assisted peak.

\begin{figure}[thb!]
	\centering
		\includegraphics[scale=0.62]{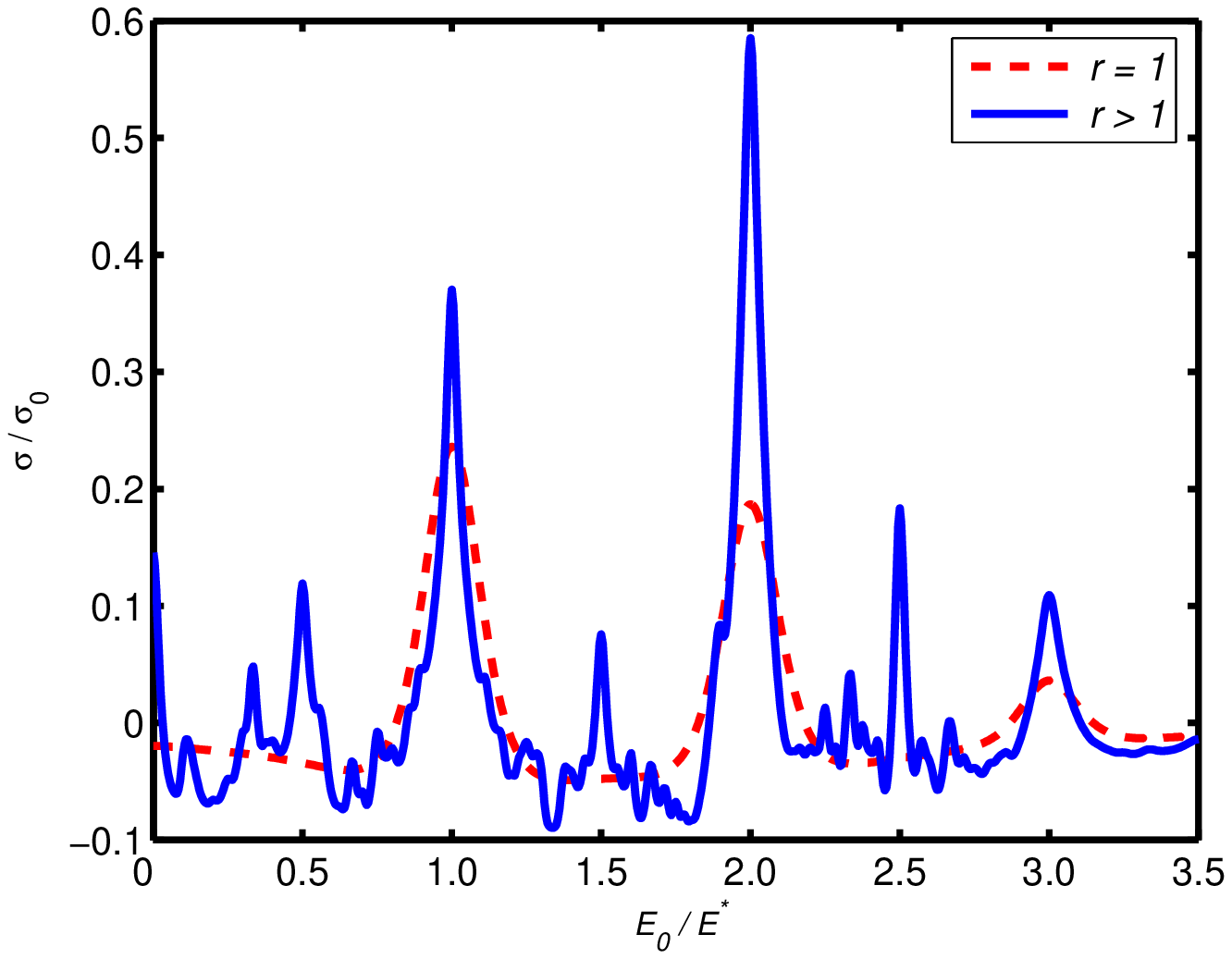}
	\caption{Fractional photon assisted process in graphene nanoribbon.}
	\label{fig:frac_integrpeaks}
\end{figure}

A very important phenomenon of fractional photon assisted processes is obtained in graphene nanoribbon when the stark component $r > 1$, even though the integer photon assisted peaks are still more pronounced. In Fig.\ref{fig:frac_integrpeaks} fractional peaks are formed at $n/r = 0.5, 1.0, 1.5, 2.0, \ldots$.

\section{Conclusion} \label{sec:Conclusions}
We have used the complete tight-binding spectrum of graphene to show the phenomenon of photon assisted processes and dynamic localization in graphene. A signature which may be responsible for rectification and amplification of ac fields. We found yet another quantum mechanical behavior of graphene nanoribbons, fractional photon assisted process. We have suggested the use of our theoretical approach for studying terahertz generation and small signal amplification in graphene nanoelectronic devices.

\end{document}